\documentclass[aps,prb,amsmath,amssymb 12 pt, singlecolumn,showpacs]{revtex4}%
\usepackage{amsmath}
\usepackage{amsfonts}
\usepackage{amssymb}
\usepackage{graphicx}%
\usepackage{natbib}
\begin{document}
\title{Nonreciprocal localization of ultrasound in a viscous medium with asymmetric scatterers}

\author{Jyotsna Dhillon$^1$}
\author{Andrey Bozhko$^1$}
\author{Ezekiel Walker$^2$}
\author{Arup Neogi$^1$}
\author{Arkadii Krokhin$^1$}

\affiliation{$^1$Department of Physics, University of North Texas, 1155 Union Circle $\#311277$, Denton, TX 76203}
\affiliation{$^2$Echonovus Inc., 1800 S Loop 288, Denton, TX 76205}

\date{\today }

\begin{abstract}
A two-dimensional phononic crystal with asymmetric scatterers is used for the study of Anderson localization of sound along one-dimensional disorder produced by random orientation of metallic rods. An exponentially weak transmission of ultrasound is demonstrated for the waves propagating along the direction of disorder. In the perpendicular direction where the scatterers are ordered, sound propagates as extended wave. The {\it PT}-symmetry of the system is broken by dissipative viscous losses and asymmetric shape of the scatterers. Nonreciprocal transmission of sound is observed for both, ordered and disordered, directions.  In the localized regime, the nonreciprocity is manifested through different values of localization length for sound propagating in the opposite directions.

\end{abstract}
\pacs{43.20.Fn, 43.20.+g, 43.30.Ft, }

\email{arkady@unt.edu}
\maketitle

\section{Introduction}
It is commonly accepted that nonreciprocal propagation of sound requires presence of nonlinear elements along the path of sound wave or broken time-reversal symmetry ({\it T}-symmetry). Both types of acoustic systems have been successfully fabricated in the recent years, see, e.g., the reviews \cite{Fle,Cumm3}. In active acoustic devices, the {\it T}-symmetry is usually broken due to rotation of the background fluid that, in the case of local vortexes, becomes topologically equivalent to presence of external magnetic field \cite{Yang} and, in the case of global rotation, leads to Doppler shift which serves as a source of nonreciprocity \cite{Hab}.

Dissipative losses  of elastic materials is not considered a nonreciprocal factor since the most common way to introduce dissipation --  adding an imaginary part to elastic coefficient -- does not break {\it T}-symmetry. This phenomenological approach conserves the reciprocity as a general property of the wave equation \cite{Hoop}. Unlike this, the wave equation obtained in the microscopic approach based on the Navier-Stokes equation is not time-reversible. A sound wave passing through a set of {\it asymmetric} scatterers, depending on direction of propagation, generates different spatial distributions of fluid velocities that may lead to different dissipative losses. The difference in viscous losses is manifested as nonreciprocity in sound propagation through a system where {\it P}- and {\it T}-symmetries are simultaneously broken \cite{Walk}. Nonreciprocity associated with energy losses was also observed in transmission of sound through a metasurface with inhomogeneous index of refraction \cite{Cumm4}. In the experiments \cite{Walk,Cumm4} nonreciprocity was observed for monochromatic plane waves. Numerical simulations confirm nonreciprocity for quasi-point emitter and receiver \cite{Walk}. Rigorous mathematical proof of reciprocity/nonreciprocity for Navier-Stokes equation is still lacking \cite{Pierce}.

Here we report our experimental and numerical results on nonreciprocal propagation of sound through a disordered set of asymmetric scatterers. Disorder, depending on its strength and dimensionality of the system of scatterers may lead to Anderson localization of waves. Destructive interference of scattered and incoming waves results in exponential decay of the incoming signal. If at a given frequency $\omega$ the length of the sample $L$ exceeds the localization length $l(\omega)$,  the transmitted signal $\sim \exp(-L/l)$ becomes so weak that the disordered medium can be considered  nontransparent (insulator). It is well-known that for a one-dimensional (1D) and two-dimensional (2D) system any weak disorder localizes  the incoming wave \cite{MT}. For 2D disordered systems, the localization length grows exponentially with the mean free path making experimental observation of localized states practically impossible since it requires an enormously large, weakly disordered system. Localization in 3D systems requires the strength of disorder exceed some critical value, at which point, a metal-insulator transition becomes possible. For light  the effect of Anderson localization was observed for different types of disorder in 1D (and quasi 1D) \cite{Chab,Lah,Sza}, 2D \cite{Rib,Stu}, and 3D systems \cite{Wie}. Modern reviews on the current state of the problem of localization of light can be found in Refs. [\onlinecite{Wie2,Seg}].

There are much less publications reporting localization of sound. Some signatures of localization of ultrasound in a polymer elastic disk subjected to liquid-solid temperature transition were reported in Ref. [\onlinecite{Gra}]. The authors measured signal attenuation and change of the sound phase velocity in a sample where random scatterers are the nucleation centers formed in the process of solidification. Approximately at the same time the localization of elastic vibrations of an aluminum plate with randomly distributed slits was observed \cite{Wea}.
An extended study of localization of sound waves in a network of elastic spheres was done in Ref. [\onlinecite{Hu}]. Several signatures of localization have been reported, including the comparison of microstructure of the localized and extended states and results of observation of transverse Anderson localization in 3D. Later on one more experiment on transverse localization was performed using a multichannel waveguide with random coupling between discrete acoustic channels \cite{Ye}.

\begin{figure}
\includegraphics [width=16cm]{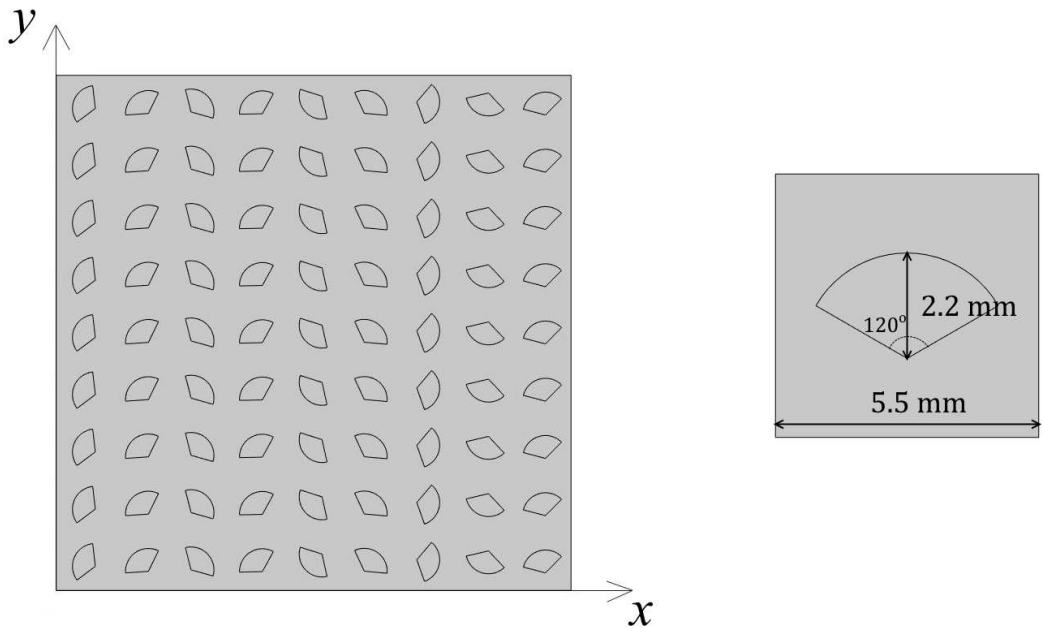}\caption{Disordered phononic crystal of $9\times9$ aluminum rods used in the experiemnts on sound transmission. Along the $x$-axis the rods are completely disoriented producing strong disorder. Right panel shows the unit cell.}
\label{fig1}
\end{figure}

Here we propose to use a phononic crystal with randomly oriented asymmetric aluminum rods imbedded in viscous liquid (water) to observe localization of ultrasound. While this arrangement of scatterers is two-dimensional it is easy to obtain one-dimensional disorder if all the rods are equally oriented within the columns but randomized along the rows, see Fig. \ref{fig1}. Due to the difference symmetry this arrangement of scatterers behaves like a crystal if sound propagates along the columns and as a 1D disordered structure if sound propagates along the rows. Transmission through a two-dimensional elastic phononic crystal with one-dimensional disorder was numerically analyzed in Ref. [\onlinecite{Yan}]. Random fluctuations were introduced through random perturbation in the filling fraction along the direction of disorder. The shape and orientation of the square scatterers was not perturbed.  Localization of elastic waves was reported. Since the unperturbed and perturbed structures do possess {\it P}-symmetry the reciprocity in transmission is conserved.     In our present study, the nonreciprocity appears as a result of broken {\it PT}-symmetry \cite{Walk}. The nonreciprocity is estimated from the transmission spectra for the waves propagating in opposite directions. These measurements are performed for two directions of propagation,  along the columns and along the rows. For propagation along disorder (rows) the localization length exhibits a difference which serves as a measure for nonreciprocity in a disordered system. Experimental results are confirmed by numerical simulations.

\section{Analysis of transmission spectra}
The experiments on sound transmission were performed using the sample shown in Fig. \ref{fig1}. It contains $9 \times 9$ aluminum rods arranged in a square lattice. Each rod has the same cross-section in a form of a circular sector with angle of $120^{\circ}$. The rods are equally oriented within the columns and disoriented along the rows. Either the number of columns or the number of rows was gradually reduced from 9 to 1 in the experiments where transmission was measured as a function of the sample length.

\begin{figure}
\includegraphics [width=16cm]{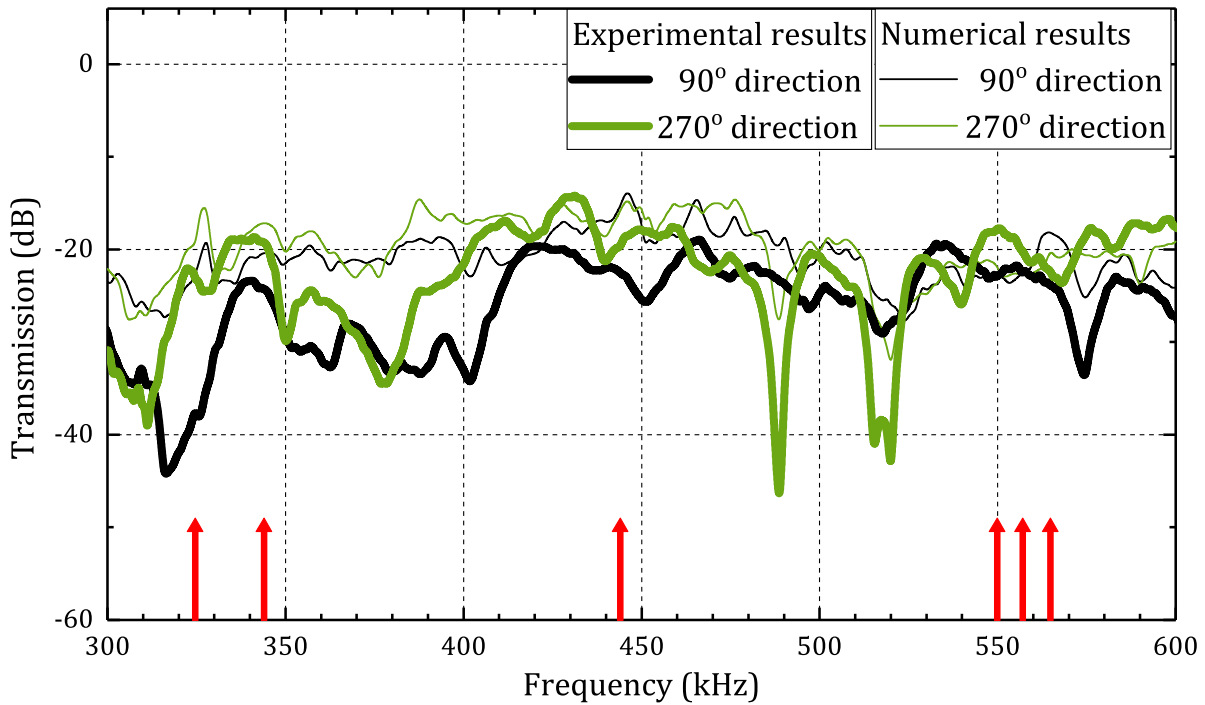}\caption{ The transmission spectra of the $6\times 9$ phononic crystal for sound waves propagating along axis $y$ (along ordered structure of 6 rows) in two opposite directions,  corresponding to the angles of $90^{\circ}$ and $270^{\circ}$. The angles are counted from the positive direction of axis $x$ in Fig. \ref{fig1}. Red vertical arrows show the resonant frequencies of a single aluminum rod in water.}
\label{fig2}
\end{figure}

\begin{figure}
\includegraphics [width=16cm]{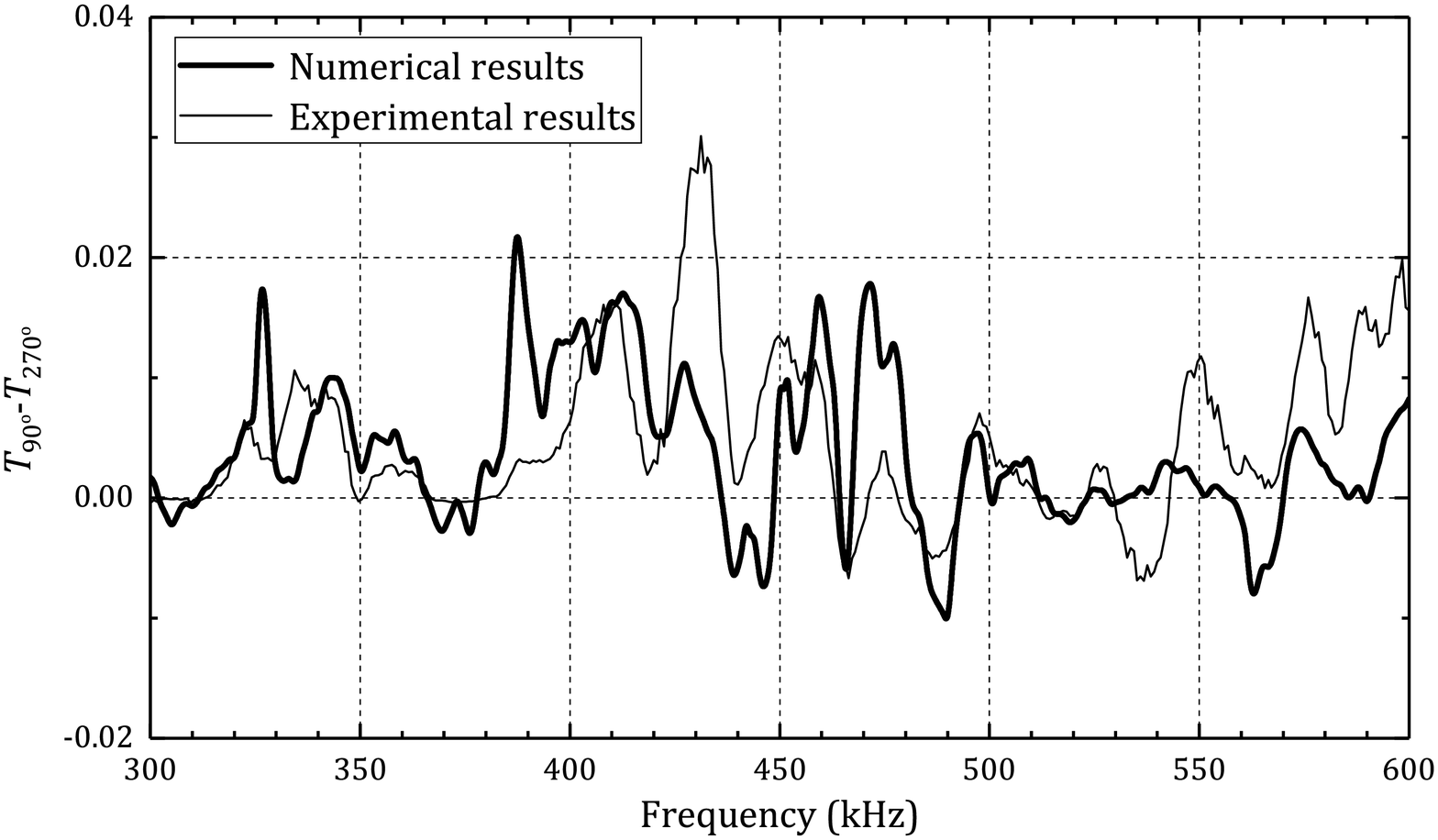}\caption{ The nonreciprocity obtained from the transmission spectra shown in Fig. \ref{fig2}. }
\label{fig3}
\end{figure}

Transmission spectra was recorded along the $x$-(disordered) and $y$-(ordered) directions using two, identical, Olympus V301 0.5 MHz $1^{”}$ unfocused immersion transducers in a bi-static, thru-transmission setup. Measurement was performed by centering the sample between the two transducers, with one transducer connected to a signal generator, and the other connected to a Tektronix MDO 3024b Spectrum Analyzer. Spectra for both the forward and reverse directions was obtained by sweeping the function generator in a continuous wave mode over the full bandwidth of the transducers at a constant input voltage, then the leads for the emitter and detector switched without perturbing the setup and the process repeated.
To obtain the transmission spectral behavior as dependent on distance, spectra was recorded for 9-1 layers (e.g. $9\times9$, $9\times8$, ..., $9\times1$) along the direction of propagation in both the $x$- and $y$- directions. For each representation of the sample, transmission was recorded with both the sample present and the sample removed for a baseline. The difference between the signals received for the sample and the baseline measured in dB was used for analysis. All measurements were performed at room temperature in a tank filled with deionized water as the ambient medium for the sample.
\newline{\it Spectra for sound propagating along the direction of periodicity.} The structure in Fig. \ref{fig1} possesses translational symmetry in the vertical direction (along axis $y$). This symmetry, however, is not the same as the symmetry of a crystal periodic in both direction. The Bloch theorem for translation by period $a$ along $y$ states that pressure $p(x,y+a)=e^{ik(x)a} p(x,y)$. Here the Bloch vector $k(x)$ is not the same for all the columns but it changes with $x$-coordinate. Therefore, the propagating monochromatic wave, $p \sim e^{-i\omega t}$, cannot be characterized by dispersion relation of a standard type, $\omega=\omega(k)$. Due to disorientation of the rods, diffraction at a set of scatterers does not lead to a regular pattern typical for a crystal, resulting in the transmission spectra in Fig. \ref{fig2} exhibiting irregular sequence of passing bands and gaps. There are pronounced minima in the transmission spectra lying near the resonances of a single rod in water. These resonances are shown by red vertical lines. Similar feature in the transmission spectrum was reported in Ref. \cite{Hu}. Other minima and maxima originate from multiple scattering and their positions are irregular. The experimental graph reproduces most of the main features predicted numerically for the opposite directions of propagation. While the nonreciprocity is relatively small (because of low viscosity of water),  it is well manifested throughout the range of frequencies. To make the nonreciprocity more visible, we plot in Fig. \ref{fig3} the difference in transmission in both directions $T_{90^{\circ}}-T_{270^{\circ}}$ versus frequency. There is a good agreement between the experimental and theoretical data on nonreciprocity. The origin of the nonreciprocity is related to the mechanism of gradient induced differential dissipation (GIDD) \cite{Walk}. This mechanism requires strongly broken {\it P}-symmetry that leads to different dissipative losses for the waves propagating in opposite directions. Along the direction of periodicity {\it P}-symmetry is essentially broken for each column of scatterers shown in Fig. \ref{fig1}.
\begin{figure}
\includegraphics [width=16cm]{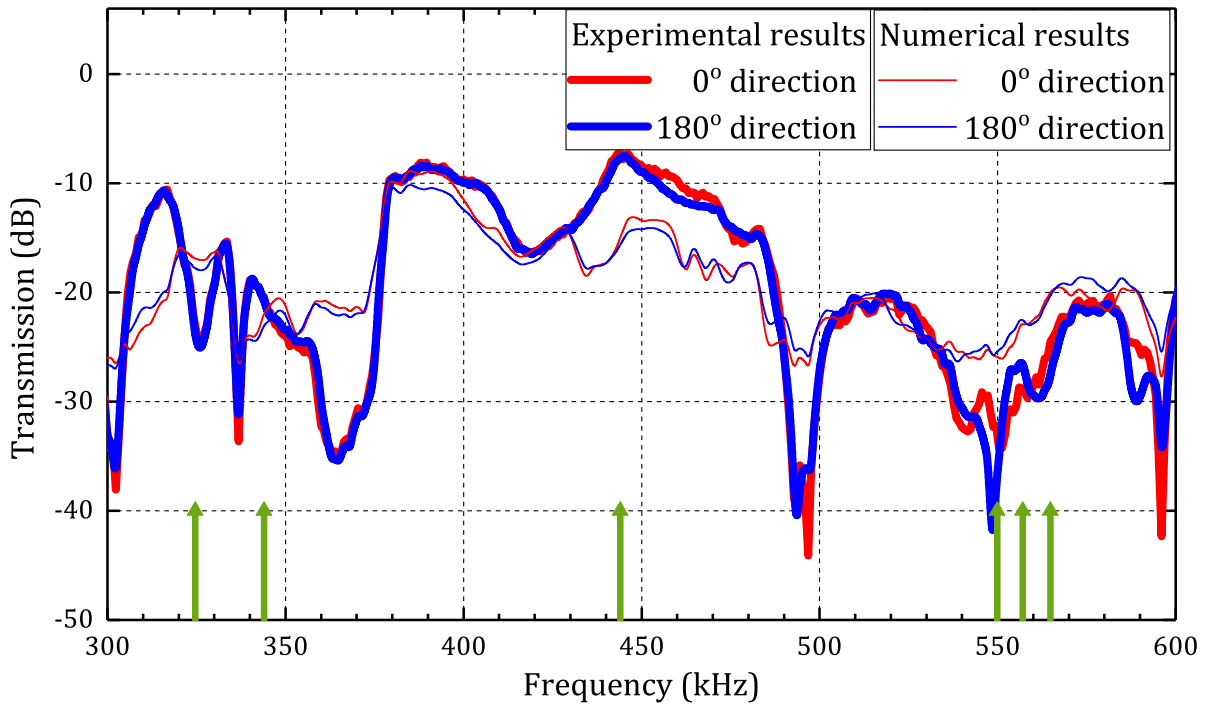}\caption{ The  transmission spectra of $9\times 6$ phononic crystal for sound wave propagating along axis $x$ (along disordered structure of 6 columns) in two opposite directions of $0^{\circ}$ and $180^{\circ}$. }
\label{fig4}
\end{figure}
\begin{figure}
\includegraphics [width=16cm]{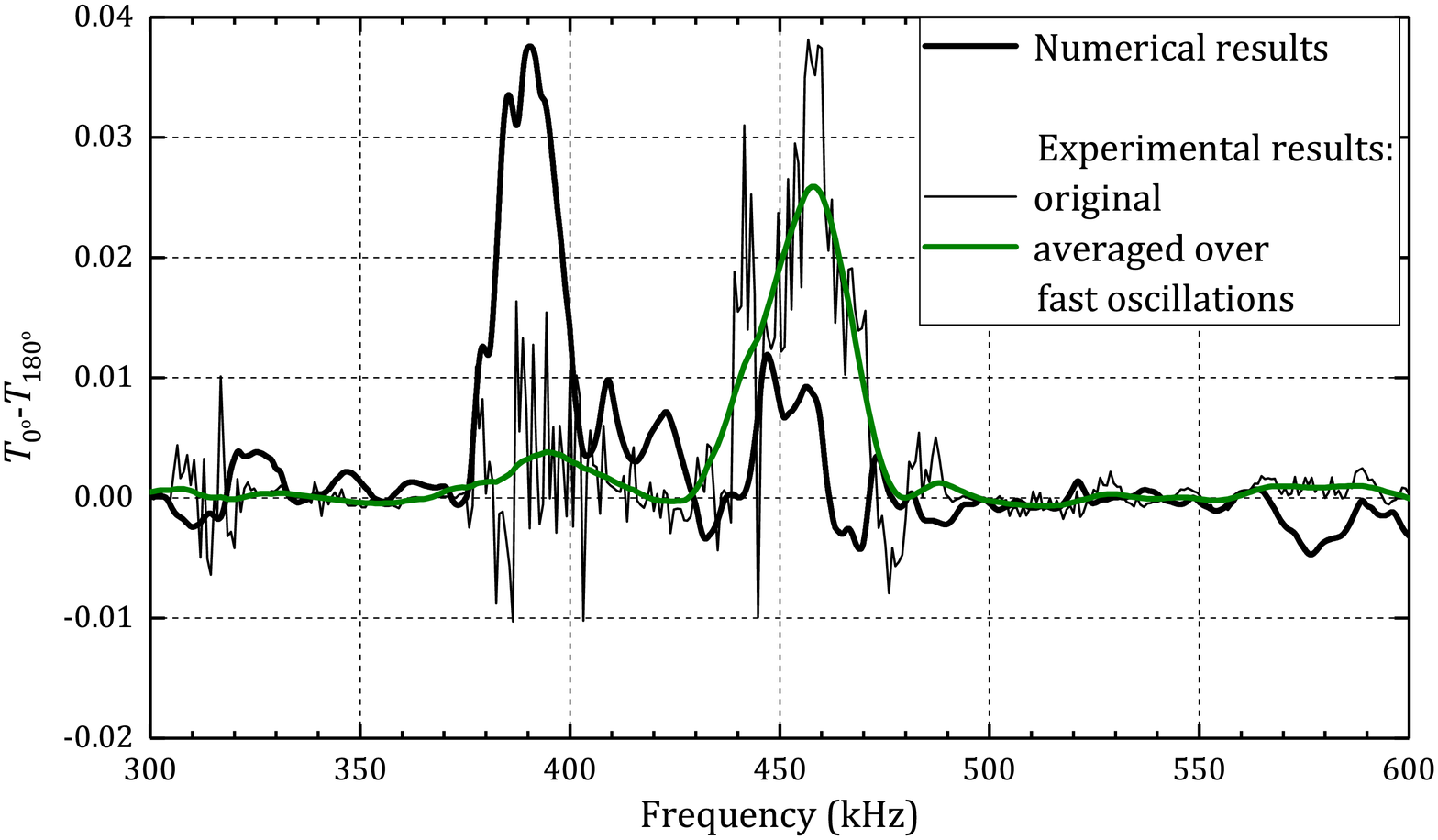}\caption{ The nonreciprocity obtained from the transmission spectra shown in Fig. \ref{fig4}. }
\label{fig5}
\end{figure}
\newline {\it Spectra for sound propagating along the direction of disorder.} Sound wave propagating along axis $x$ suffers from strong random scattering and the transmission spectra in Fig. \ref{fig4} do not show any regular structure. The nonreciprocity is clearly visible, however in this case two spectra overlap for relatively wide region of frequencies. The nonreciprocity plot in Fig. \ref{fig5} is less than that in Fig. \ref{fig3} for almost all frequencies. The reduction of nonreciprocity is due to higher {\it P}-symmetry observed for strongly disordered set of scatterers. Indeed, the scatterers in each row in Fig. \ref{fig1} are completely disoriented and the number of scatterers oriented in any direction is approximately the same. An infinite chain of strongly disordered scatterers is statistically {\it P}-symmetric and does not exhibit nonreciprocity. For a finite-length  sample {\it P}-symmetry is broken only due to statistical fluctuations, which decay as $1/\sqrt L$ with length of the sample. Our results for nonreciprocity in sound transmission through longer samples, of 7, 8, and 9 columns, show gradual decrease of nonreciprocity. Transmission coefficient of sound propagating along the direction of disorder exhibits fast oscillations with frequency. The experimental results for nonreciprocity in Fig. \ref{fig5} obtained by averaging over these oscillations.

\section{Localization length}
\begin{figure}
\includegraphics [width=16cm]{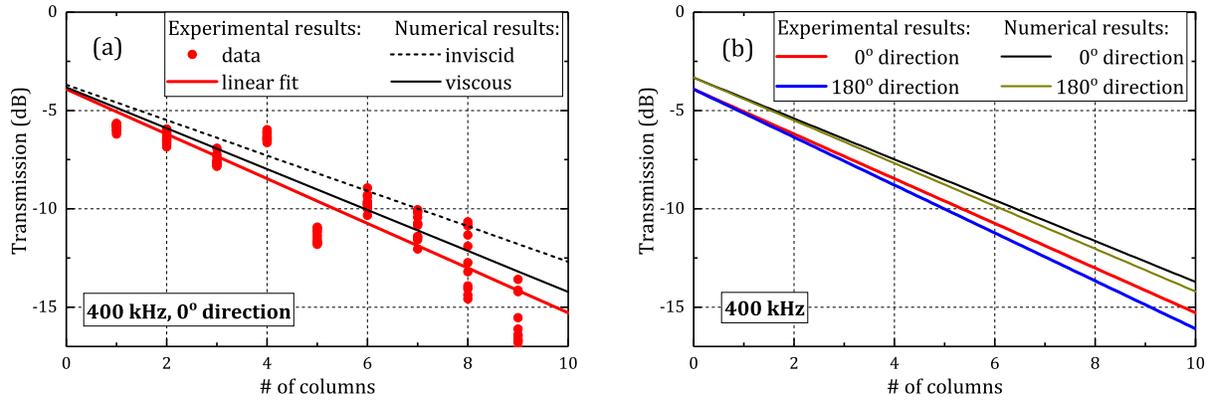}\caption{ Transmission coefficient (in dB) along the direction of disorder (axis $x$) as a function of the sample length. Slope of the linear fit is proportional to the inverse localization length $l^{-1}$. (a) Comparison of experimental and numerical results obtained for the wave propagating along $0^{\circ}$ direction. Dashed line obtained numerically for ideal (inviscid) water shows that the exponential decay of the transmission with sample size is mostly due to the effect of Anderson localization. (b) Linear fit of the transmission coefficient obtained for two opposite direction of propagation ($0^{\circ}$ and $180^{\circ}$). Lines in (b) are offset vertically for clarity. }
\label{fig6}
\end{figure}

Anderson localization of sound propagating along the direction of disorder was in the experiment with the samples of different length. The localization length is calculated as
\begin{equation}
l^{-1}(f)=\frac{2}{L} \langle\ln \frac{I_0}{I(L,f)}\rangle,
\label{1}
\end{equation}
where $I(L,f)$ is the intensity of transmitted sound and averaging is taken over statistical ensemble of disordered samples. In the experiment and numerical simulations, the latter averaging is replaced by averaging over 11 frequencies lying within a symmetric narrow window of 8 kHz centered at frequency $f$. The linear fit of the average logarithm of the transmission coefficient versus sample length is plotted in Fig. \ref{fig6} for frequency $f=400$ kHz. In the vicinity of this frequency the transmission coefficient is a relatively flat function that allows one to replace statistical averaging by averaging over frequency. Linear decay of the average transmission (in dB) is a signature of Anderson localization. The slope of the lines in Fig. \ref{fig6} gives the following values for the localization length measured in lattice period (5.5 mm). For sound propagating along $0^{\circ}$-direction  $l_{exp}=7.64$, $l_{num}=8.36$. For sound propagating along $180^{\circ}$-direction  $l_{exp}=7.13$, $l_{num}=7.99$. Experimental results give stronger localization since there are some factors such as surface roughness of the rods and radiation losses that lead to faster decay of the signal experimentally, but are ignored in the theory.

Viscous dissipation behaves as an additive to the Anderson localization contribution of the exponential decay of the wave propagating in a disordered medium. Therefore, in many cases it is hard to separate these two contributions. Moreover, dissipation in an electronic system means inevitable inelastic scattering that is a source of phase-breaking processes that destroy localization. The situation is better for classical waves since dissipation usually reduces the amplitude of scattered wave, leaving its phase unchanged. Nevertheless, dissipation may strongly suppress Anderson localization or mislead the interpretation of experimental results\cite{John,PWA}. The original method of detection of Anderson localization in a medium with optical absorption based on analysis of electromagnetic fluctuations in transmitting channels was proposed in Ref. [\onlinecite{Chab}].

While dissipation plays a principal role in our experiments with nonreciprocal propagation of sound, its contribution to the decay of sound is quite small as compared to backscattering, which is the physical reason of localization. In Fig. \ref{fig6}a the dashed line shows the numerical results for transmission versus sample length in ideal inviscid water. It is clear that the slope of the dashed line is very close to the slope of the lines obtained for viscous water. The decay of sound in inviscid water is solely due to Anderson localization. The slope of the dashed line gives the localization length $l_{ideal} = 9.66$ which is only $16\%$ longer than that calculated for viscous water. Thus, the level of dissipation in water is not sufficient to essentially suppress Anderson localization, but it turns out to be sufficient for observation of nonreciprocal transmission of sound along the direction of order and disorder in 2D phononic crystal.

It is interesting to compare the decay of transmission coefficient along axis $x$ (disorder) and along axis $y$ (order). While the wave propagating along axis $y$ is not a standard Bloch wave, it carries information about translational periodicity and does not decay exponentially. Numerical simulation of the decay of the transmission coefficient with sample length clearly show that sound intensity decays much slower than exponential law along $y$ (black triangles) and has a tendency to saturation. Thus, we may conclude that sound propagating along axis $y$ is represented by extended states. These states may exhibit some signatures of wave chaos but this question requires a detailed study of distribution of resonances within much wider region of frequencies.

The phononic crystal used in the experiments is intentionally designed to have the maximal level of disorder. This is done in order to have localization length shorter than the length of the sample. Experiments with weakly disordered phononic crystals require much longer samples but they are expected to demonstrate stronger non-reciprocity. As it was mentioned before, a strongly disordered sample is statistically {\it P}-symmetric that leads to relatively weak nonreciprocity. In a weakly disordered sample all the scatterers are oriented approximately in the same direction, thus providing strong violation of {\it P}-symmetry.  The measure of disorder in a sample shown in Fig. \ref{fig1} is the root-mean-square $\sigma = \sqrt{\langle \phi_n^2 }\rangle$ of the angle $\phi_n$ fluctuating with the column number $n$. In an ordered crystal, all the scatterers are oriented along the direction $\phi_n=0$. In a weakly disordered sample, $\sigma \ll 1$, localization occurs at long distances. According to Thouless\cite{Thoul} the localization length scales with $\sigma$ as $l \propto \sigma^{-2}$. Lack of sufficiently long samples makes the observation of localization in weakly disordered samples problematic. One more reason to study localization in weakly disordered phononic crystal is the potential to observe a mobility edge. Appropriate correlations in a random sequence $\phi_n$ may suppress localization and give rise to the mobility edge in the transmission spectrum\cite{Izr}. Correlations-induced mobility edges were observed in the microwave  transmission spectra of a metallic waveguide with periodically distributed fluctuating-strength scatterers\cite{Kuhl}. The interval of frequencies between two mobility edges where all the states are localized can be considered as a mobility gap as it was recently proposed in Ref. [\onlinecite{Page2}], where such anomaly in the spectrum of elastic vibrations of strongly disordered 3D sample has been observed. It is clear that a mobility gap may also be designed for 1D weakly disordered phononic sample of sufficient length by introducing appropriate correlations to the sequence of random angles $\phi_n$.

\section{Conclusions}

We propose a new type of disordered phononic crystal where asymmetric solid rods are periodically arranged in 2D lattice and the deliberately and controlled disorder is introduced due to random orientation of each rod. This design allows fabrication of 2D disorder, as well as of 1D disorder where the  rods of the same column are equally oriented but the columns are disoriented with respect to each other. Analysis of the transmission spectra of the phononic crystals of different length leads to the conclusion that sound waves propagating along disordered direction are exponentially localized but the waves running in the perpendicular direction (along periodicity) are extended, although they are not associated with standard Bloch waves in 2D lattice. Since the phononic crystals are imbedded in a viscous water the {\it T}-symmetry is broken and the {\it P}-symmetry is broken due to asymmetry of the scatterers. As a result of broken {\it PT}-symmetry the transmission spectra along the opposite directions exhibit nonreciprocity. Experimental and numerical results for the transmission and nonreciprocity are in a reasonable agreement. The proposed 2D phononic crystal with 1D disorder is a strongly anisotropic metamaterial with metallic behavior along direction of order and insulating behavior along direction of disorder. The results of metal-insulator transition with respect to the direction of propagation of sound wave will be published elsewhere.

\section{Acknowledgements}
This work is supported by an
Emerging Frontiers in Research and Innovation grant from
the National Science Foundation (Grant No. 1741677). The
support from the Advanced Materials and Manufacturing
Processes Institute (AMMPI) at the University of North
Texas is gratefully acknowledged.

\begin{figure}
\includegraphics [width=16cm]{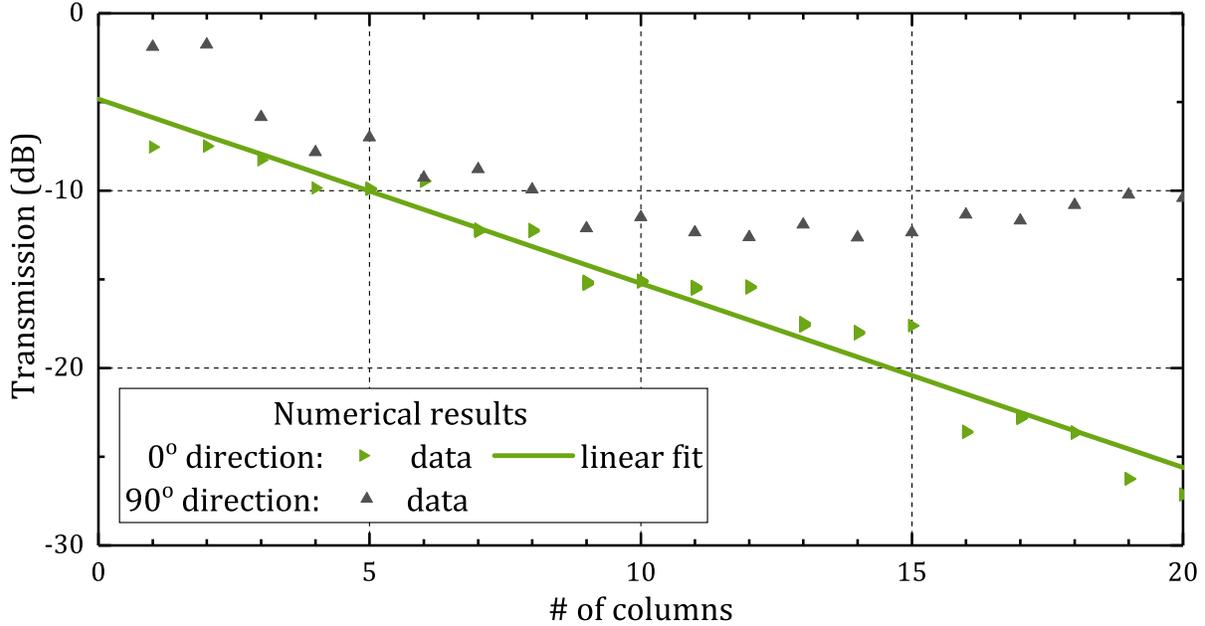}\caption{ Numerically calculated transmission coefficient (in dB) along the direction of disorder ($0^{\circ}$, green trangles) and along the direction of order ($90^{\circ}$, black triangles) as a function of the sample length. Nonlinear decay along axis $y$ is a clear evidence that the corresponding eigenstates are extended.}
\label{fig7}
\end{figure}

\end{document}